\documentstyle[epsfig,12pt]{article}

\newcommand{\beq}{\begin{equation}}
\newcommand{\eeq}[1]{\label{#1} \end{equation}}
\newcommand{\beqar}{\begin{eqnarray}}
\newcommand{\eeqar}[1]{\label{#1} \end{eqnarray}}

\title{\begin{flushright}
{\normalsize NUC-MINN-2001/18-T}
\end{flushright}
\vspace*{0.4in}
{\bf Domain Wall Dynamics of Phase Interfaces}}
\author{L. P. Csernai$^{1,2,3}$,  J. I. Kapusta$^3$ and E. Osnes$^4$ 
\vspace*{0.2in}\\
$^1$ {\it Section for Theoretical and Computational Physics,
Department of Physics}\\  \vspace*{0.1in} {\it University of
Bergen, Allegaten 55, 5007 Bergen, Norway}\\
$^2$ {\it KFKI Research Institute for Particle and Nuclear
Physics}\\  \vspace*{0.1in} {\it P.O.Box 49, 1525 Budapest, Hungary}\\
$^3$ {\it School of Physics and Astronomy, University of Minnesota}\\
\vspace*{0.1in}
{\it 116 Church Street SE, Minneapolis, MN 55455, USA}\\
$^4$ {\it Department of Physics, University of
Oslo}\\ {\it P.O.Box 1048 Blindern, 0316 Oslo 3, Norway}}

\date{}

\begin{document}

\maketitle

\begin{abstract}
The statics and dynamics of a surface separating two phases of a relativistic 
quantum field theory at or near the critical temperature typically make use of a 
free energy as a functional of an order parameter.  This free energy functional 
also affords an economical description of states away from equilibrium.  The 
similarities and differences between using a scalar field as the order parameter 
versus the energy density are examined, and a peculiarity is noted.  We also 
point out several conceptual errors in the literature dealing with the dynamical 
prefactor in the nucleation rate.

\end{abstract}

\newpage

\section{Introduction}

Different phases of matter are separated in space and time by dividing 
layers called domain walls. The dynamics of the domain walls is the 
determining mechanism governing phase transitions in an intermediate
range between very slow, quasi-static and very rapid, dynamical processes.
In this intermediate range the phase transition speed and the speed of
external constraints are comparable to each other.

The dynamics of phase transitions is an involved subject even in macroscopic
systems. First of all, phase transitions can be different. They
may include slow burning or deflagration, detonation,
condensation, evaporation, and many other forms of transition. Nevertheless,
the basic conditions of all these transitions have some
similarities.  These arise from the basic conservation laws and from the
requirement of local, or at least approximately local, equilibrium.

In a dynamical situation the approach using the equation of state with a first
order phase transition is identical both in compression and in expansion.  
If the compression is supersonic, shock
or detonation waves are formed where the final new phase is immediately
created. The phase transition speed influences only the width of the
shock front. On the other hand, for slow dynamics and rapid phase transition the
shock front width is primarily determined by the transport coefficients,
viscosity and heat conductivity, and not by the phase transition speed.

At high energies a relativistic treatment is frequently necessary.
It is important to mention that a system is also relativistic if the matter is
radiation dominated, meaning that the rest mass of the constituent particles
is zero or negligible compared to the fourth root of the energy density.
These types of systems must be treated as relativistic even if the collective 
velocities are small.
This was one of the important new features recognized in ref. \cite{CK92D}.
If conserved charges do not exist the application of
the conventional theory of phase transition dynamics \cite{LT73,TL80}
is not possible, and this leads to essential differences in the phase
transition dynamics. Most importantly the flow is tied to
the energy flow; thus energy or heat current and the fluid flow are identical
and heat conduction (with respect to the flow) may not occur. Therefore
the coefficient of heat conductivity does not exist, and cannot govern
the phase transition speed.

The spatial configuration of phase transition instabilities varies in
large homogeneous systems. When the conditions for the occurrence of a
new phase are established the most common form of the appearance of the
new phase happens via the formation of small critical size bubbles or
droplets. (Sub-critical size bubbles or droplets will shrink and vanish
but supra-critical size ones will grow.) This configuration is called
homogeneous nucleation.

As the level of supercooling increases and the amount of the new phase
increases other geometries become energetically more favorable, such as
elongated cylindrical objects (spagetti) or layers (lasagna). The phase 
transition dynamics is then referred to as spinodal decomposition, indicating
that systems which supercool (or superheat) and reach the adiabatic or
isothermal spinodals on the phase diagram start the formation
of the new phase in these configurations immediately.

Finally, when the two phases are about equally abundant and/or the
transition is extremely rapid the two phases form a somewhat
random occupation of the configuration space called percolation.

Most frequently explicit dynamical calculations are performed for the
homogeneous nucleation geometry as this is usually the initial and the
slowest of all.  Nevertheless, the domain walls and their dynamical properties 
play an important role in all of the above mentioned configurations.
One important feature of the nucleation studies that is not always recognized
is that within the dynamical domain wall or droplet wall we do not have
thermal equilibrium.  Here, quantities like energy, entropy or particle density 
make sense, while other quantities that rely on thermal equilibrium, such as the 
equation of state, may not.

If the fluid is perfect and consists of a
single phase of matter in full phase equilibrium, the flow is adiabatic
\cite{Cs94-257}.  Nevertheless entropy production is still possible if the 
system is out of phase equilibrium \cite{Cs94-257}, or out
of thermal equilibrium. Thus, in domain walls entropy may be produced
even in perfect fluids as here we do not have an equation of state
which would satisfy the requirements of equilibrium thermodynamics.

In this paper we would like to discuss connections between the two most 
frequently encountered approaches to phase transitions using 
effective field theoretical models and 
phenomenological thermal and fluid dynamical methods.  In section 2 we 
first demonstrate with a simple effective field theoretical model how one can 
obtain a Landau-model type of phenomenological 
description of critical dynamics. In section 3 we 
elaborate on the assumptions and error of different approaches used 
recently in the literature. We present a summary and conclusion in section 4.

\section{The Free Energy Functional}

A common approach to dealing with a phase transition in relativistic quantum 
field theory is to assume a uniform condensate $\phi$ for some scalar field and 
then to compute quantum and thermal fluctuations about that condensate
\cite{Kapbook}.  This is 
especially so for systems exhibiting spontaneous symmetry breaking where the 
condensate field serves as an order parameter to distinquish the two phases.  
Examples include the Higgs field in electroweak theory and the sigma field or 
scalar quark condensate in strong interaction physics.  The resulting free 
energy density $f$ is obtained from the partition 
function $Z$ in the usual way.
\begin{equation}
f(\phi,T) = -T \ln Z(\phi,T,{\cal V})/{\cal V}
\end{equation}
Here ${\cal V}$ is the volume.  Since the high temperature symmetric phase 
usually corresponds to $\phi = 0$, it is customary to define the effective 
potential as the deviation of the free energy from its value in the symmetric 
phase.
\begin{equation}
V(\phi,T) = f(\phi,T) - f(0,T)
\end{equation}
The point $\phi = 0$ is either a global minimum or only a local minimum 
depending on whether $T$ is greater than or less than the critical temperature 
$T_c$, respectively.  Therefore the results of computations are often found to 
be, or parametrized as, a fourth order polynomial in $\phi$.
\begin{equation}
V(\phi,T) = \sum_{n=2}^4 a_n(T) \phi^n
\end{equation}
Occasionally one will find additional terms of order $\phi^4 
\ln\left(\phi/\Lambda\right)$ or higher powers of $\phi$, but these are not 
common and will not change our analysis qualitatively.

To give a specific example we will use the parametrization of ref. \cite{EI92}.  
\begin{equation}
V(\phi,T) = \frac{1}{2} \gamma (T^2-T_0^2) \phi^2
-\frac{1}{3} \alpha T \phi^3 + \frac{1}{4} \lambda \phi^4
\end{equation}
The $\gamma$, $T_0^2$, $\alpha$ and $\lambda$ are temperature independent 
constants, to be specified shortly.  In an equilibrium state the free energy 
density is the negative of the pressure,
$f(T)=-p(T)$.  Normalizing to the high temperature 
symmetric phase, and denoting the equilibrium pressure of that phase by 
$p_h(T)$, we have
\begin{equation}
f(\phi,T) = -p_h(T) + V(\phi,T) \, .
\end{equation}
It is not enough to specify the effective potential; 
the equilibrium pressure as 
a function of temperature must be specified too.

If one knows the free energy as a function of $T$ then the energy density $e$ 
can be calculated from the basic thermodynamic identities.  
In particular, if we 
want to know the energy density for a specific value of the condensate 
field, including out of equilibrium configurations also, 
then we must hold it fixed during the temperature differentiation. 
\begin{eqnarray}
e(\phi,T) &=& f(\phi,T) - T \frac{\partial}{\partial T}f(\phi,T)\nonumber \\
&=& e_h(T) - \frac{1}{2} \gamma (T^2+T_0^2) \phi^2
+ \frac{1}{4} \lambda \phi^4
\end{eqnarray}
The first line above is the general thermodynamic identity; the second line 
applies to the specific parametrization under discussion.  If one is at a local 
minimum of the effective potential, either $\phi_h(T) = 0$ or $\phi_l(T) > 0$ 
corresponding to the high and low temperature phases, respectively, then it 
doesn't matter whether the order parameter is held fixed during the 
differentiation or not because $\partial f(\phi,T)/\partial \phi = 0$ at those 
points.  These two points correspond to thermodynamical equilibrium.

If the system undergoes a first order phase transition at a 
critical temperature 
$T_c$ then at that temperature $f$ will have two degenerate minima.  For a 
finite range of temperature above $T_c$ there will persist a higher, local 
minimum at $\phi_l(T) > 0$. For a finite range of temperature below $T_c$ there 
will persist a higher, local minimum at $\phi_h(T) = 0$.  The most important 
physical quantities are the latent heat 
$L = \Delta e(T_c) = e_h(T_c)-e_l(T_c)$, 
the correlation lengths $\xi^{-2} = \partial^2 f(0,T_c)/\partial \phi^2 =
\partial^2 f(\phi_l(T_c),T_c)/\partial \phi^2$, which are equal for the fourth 
order effective potential at $T_c$, and the surface energy $\sigma$. 
At the critical temperature the planar interfacial profile has a nice 
analytical 
solution on account of the fact that the effective potential becomes symmetric.  
In the usual way one finds that the profile field $\overline{\phi}(x)$ is
\begin{equation}
\overline{\phi}(x) = \frac{\phi_l(T_c)}{2} \left[ 1 - \tanh(x/2\xi) \right] \, .
\end{equation}
This interpolates through intermediate non-equilibrium states from
$\phi_l(T_c)$ for $x \ll -\xi$ to $\phi_h(T_c) = 0$ for $x \gg \xi$.
The surface energy can then be expressed as
\begin{equation}
\sigma = \int_{-\infty}^{\infty} dx \, \left( \frac{d\overline{\phi}}
{dx} \right)^2 \, .
\end{equation} 
The parameters in the effective potential can be expressed in 
terms of these physical parameters as follows.  
\begin{eqnarray}
\gamma &=& \left( 1 + \frac{L\xi}{6\sigma}\right) \frac{1}{T_c^2\xi^2}
\nonumber \\
T_o^2 &=& \frac{L\xi}{L\xi + 6\sigma} T_c^2 \nonumber \\
\alpha &=& \frac{1}{3\sigma \xi^3} \nonumber \\
\lambda &=& \frac{1}{T_c \xi}\sqrt{\frac{3}{2\sigma \xi^3}}
\end{eqnarray}
For purposes of illustration, and with the QCD phase transition in mind, we 
choose the following numerical values.
\begin{eqnarray}
\xi &=& \frac{1}{T_c} \nonumber \\
\sigma &=& \frac{17}{6}\frac{\pi^2}{45}T_c^3 \nonumber \\
L &=& \frac{68 \pi^2}{45} T_c^4
\end{eqnarray}
For the pressure in the high temperature phase we take 37 effective massless 
bosonic degrees of freedom corresponding to gluons and two flavors of quark, 
plus a bag constant $B = L/4$ to simulate confinement.
\begin{equation}
p_h(T) = \frac{37 \pi^2}{90} T^4 - B
\end{equation}
The numerical value for the latent heat $L$ corresponds to a transition from the 
37 degrees of freedom mentioned above to 3 degrees of freedom for massless 
pions.  It is convenient to make everything dimensionless by measuring energies 
in units of $T_c$ and lengths in units of $1/T_c$, which practice we adhere to 
in the rest of this section.

The effective potential (4) as a function of $\phi$ is plotted in figure 1 for 
three 
different temperatures: the critical one and 1\% above and below it.  This 
displays the traditional behavior of a strong first order phase transition.  The 
plot includes negative values of $\phi$ which may or may not be allowable, 
depending on the origin of $\phi$ and its physical interpretation.

The free energy density (5) as a function of $\phi$ is 
plotted in figure 2 for the 
same temperatures.  The only difference between this and figure 1 is the 
addition of the pressure as evaluated in the high temperature symmetric phase.

The energy density (6) as a function of $\phi$ is 
plotted in figure 3.  It has a 
maximum at $\phi = 0$.  This follows directly from the expression for $e$ given 
earlier: For small values of $\phi$ the deviation from $e_h(T)$ is quadratic in 
$\phi$ with negative curvature.  Note that the energy density goes negative when 
$\phi$ is greater than about 2.0.

Rather than using the field $\phi$ as the order parameter 
one might consider other 
choices more appropriate for the problem at hand.  
For example, one could take as 
the order parameter the deviation of the energy density from its equilibrium 
value and expand the free energy in a power series in this difference.  This is 
the Landau approach to the description of fluctuations and to departures from 
equilibrium states \cite{Lifshitz}.  Such an approach to the nuclear 
liquid-gas phase transition was implemented by Goodman {\it et al.} 
\cite{goodman}.  It is also this same function that was used in working out the 
nucleation of relativistic first order phase transitions by two of us 
\cite{CK92D}.  With these motivations we plot the free energy versus the energy 
in figure 4 by using $\phi$ as a parameter.  There are a number of points to be 
made concerning this figure.  First, there are two minima at each of the chosen 
temperatures.  The minima occur at the energy densities $e_l(T)$ and $e_h(T)$.  
The lower of these two represents the equilibrium state while the other 
represents a metastable state except at $T_c$ when they are degenerate.  The 
value of $f$ at a minimum is equal to the negative of the pressure in that 
particular phase: $p_l(T) = -f(e_l(T),T)$ and $p_h(T) = -f(e_h(T),T)$.  Second, 
the value of $e_l(0.99T_c) < 0$!  This is an indication of the inadequacy of the 
specific parametrization of the effective potential at this temperature, and 
probably at lower temperatures too.  This is an important point to always be 
aware of when writing down any formula for an effective potential.  Third, the 
free energy has a cusp at the location of the high density symmetric phase.  
This is an unavoidable consequence of the inference of the free energy from the 
effective potential.  From figure 3 or from the corresponding formula it is 
clear that the energy density is a maximum at $\phi = 0$.  This means that the 
plot of the free energy ends at the equilibrium density $e_h(T)$ of the 
symmetric phase with a cusp.  
Physically this must be a 
restriction of fluctuations to those associated with $\phi$ and only $\phi$.  
This is too restrictive by far; there are certainly other physical processes not 
taken into account.  For example, placing particles in a box in contact with a 
heat reservoir allows for the exchange of energy between the particles and the 
reservoir, resulting in fluctuations in energy and pressure.  These processes 
are always present in systems at fixed temperature and are not accounted for, or 
associated with, fluctuations in an order parameter.

Now let us follow the Landau construction of the free energy away from 
equilibrium states using the energy density as the order parameter with no 
reference to $\phi$ whatsoever.  This is the construction made in \cite{CK92D} 
in the context of constructing spherical surfaces separating two phases at 
noncritical temperatures.  The total free energy of interaction is the sum of a 
gradient energy and a free energy density integrated over space.
\begin{equation}
F_I\{e({\bf x})\} = \int d^3x \left[\frac{1}{2}
K (\nabla e({\bf x}))^2
+f(e({\bf x}),T) \right]
\end{equation}
The $f$ is most economically written as a fourth order polynomial in $e$.
\begin{equation}
f(e,T) = \sum_{n=0}^4 b_n(T) e^n
\end{equation}
The coefficients are functions of temperature.  They are determined by several 
requirements.  The first is that $f$ has minima located at $e_l(T)$ and 
$e_h(T)$, such that $f$ at those points be equal to the negative of the 
corresponding equilibrium pressure.
\begin{eqnarray}
f(e_h(T),T) &=& -p_h(T) \nonumber \\
f(e_l(T),T) &=& -p_l(T) 
\end{eqnarray}
This results in four equations involving the five $b_n$.  Two more equations 
result from fixing the correlation length (which is the same in the two phases 
only at $T_c$) and the surface energy $\sigma$ (which, strictly speaking, is 
only well defined at $T_c$).  This set of six equations is not inconsistent 
because the coefficient of the gradient energy, $K$, must also be determined.
\begin{eqnarray}
\partial^2f/\partial e^2|_{e=e_h(T_c)} &=&
\partial^2f/\partial e^2|_{e=e_l(T_c)}
= \frac{6\sigma}{\xi L^2} \nonumber \\
K &=& \frac{6\sigma \xi}{L^2}
\end{eqnarray}
We insist upon the same equilibrium energy densities and pressures as used in 
the effective potential approach, and use the same correlation length and 
surface energy at $T_c$ too.
\begin{eqnarray}
\overline{e}(x) &=& \frac{1}{2} \left[ e_h(T_c) + e_l(T_c)
 + \Delta e(T_c) \tanh(x/2\xi) \right] \\
\sigma &=& K \int_{-\infty}^{\infty} dx \, 
\left( \frac{d\overline{e}}{dx} \right)^2
\end{eqnarray}
Then the Landau expansion of the free energy as a 
function of the energy density is obtained and is plotted in figure 5.  The 
$f(e,T)$ is now a smooth and well-behaved function.  Its only failing is that 
the equilibrium energy density at $0.99T_c$ is negative, but that is a 
consequence of insisting that it be the same as for the effective potential, 
which was negative. (Actually, fluctuations into states with negative energy can 
be avoided if one uses the Laurent expansion instead of the fourth order 
polynomial approximation \cite{CN94}, but only if the lower minimum is at 
positive energy density.)  In fact, with the Landau approach we have a closer 
connection between the physical observables and the free energy.  For example, 
we can easily specify the equilibrium energy density and pressure in each phase, 
making it simple to avoid such unwelcome behavior as a negative energy density 
in equilibrium.  Of course the direct
connection with the condensate field $\phi$ is lost.

The main difference between the two approaches is that the field theoretical 
approach parametrizes the out-of-equilibrium configurations in terms of an 
effective potential as a function of $\phi$ that shows smooth, quadratic minima 
around the two equilibrium states, whereas the Landau type of thermodynamical 
approach does the same in terms of the free energy as a function of the energy 
density $e$.  These two approaches are not fully equivalent: the
highly nonlinear $f(e)$ dependence obtained from this particular field 
theoretical model does not yield a smooth quadratic minimum in terms of $e$ at 
$e_h$ (where $\phi=0$). This leads to unphysical estimates for energy density 
fluctuations, in our opinion.  Which approach one takes depends on the physical 
situation being addressed.

\section{Dynamics of First Order Phase Transitions}

A phase transition occurs because of some change in the global properties of the 
system.  Examples of such changes are: an expansion chamber or a sudden quench 
in temperature in the laboratory, the expansion of hot matter in a high-energy 
heavy ion collision, and the expansion of the early universe.  Generally we need 
to compare the phase transition rate to the rate of expansion or quench. The 
speed at which the phase transition proceeds comes into play when the speed at 
which the external variables, such as the volume, becomes comparable to or 
exceeds that of phase conversion.

If the rate of change of an external variable is slow there is sufficient time 
to maintain phase equilibrium on a coarse-grained time scale.  This also means 
that all other equilibration processes are completed as these invariably require 
less time and less interaction than the conversion of one phase of matter to 
another.

Thus, in the case of slow external dynamics and rapid phase
equilibration the matter is in complete equilibrium, including phase
equilibrium, and the equation of state of the matter undergoing a first order 
phase transition is given by the Maxwell construction. Then we have a fully
developed mixed phase, and the phase abundances are typically given by the 
conservation of energy and entropy.  No information on dynamical processes is 
needed. This type of transition is adiabatic.
Even in moderately fast dynamical expansions there are only small deviations
from the ideal and complete phase equilibrium (Maxwell construction).
This deviation results in some delay in the creation of the new phase,
leading to supercooling or superheating and extra entropy production.
\cite{Cs94-257}

For heavy ion reactions the first attempt to explicitly evaluate the
phase transition speed of homogeneous nucleation was described
in \cite{CK92D,CK92L}. The homogeneous nucleation mechanism correctly 
describes the initial stage of the phase transition when
the abundance of the newly created phase is still small and when the
rate of phase conversion is the slowest process.

Here a couple of remarks are necessary. There are several requirements to form 
critical-sized bubbles or droplets of the new phase.  Both
pressure balance and temperature equality should be established between the
phases; this requires the transfer of energy and momentum across the phase 
boundary. If local equilibrium is assumed both before and after the formation of 
the new phase then we cannot relax the requirement of pressure and temperature 
equilibrium.

Langer's modern theory of nucleation yields the following formula for
the rate:
\begin{equation}
I = \frac{\kappa}{2\pi} \Omega_0 e^{-\Delta F/T}
\end{equation}
where $\Delta F$ is the change in the free energy of the system due to the
formation of the critical droplet. $\Omega_0$ is a statistical prefactor
which measures the available phase space volume.  $\kappa$ is a dynamical
prefactor which determines the exponential growth rate of critical droplets
which are perturbed from their quasi-equilibrium radius R$_*$.
\begin{equation}
\kappa = \frac{d}{dt}\ln [R(t) - R_*]
\end{equation}
The dynamical prefactor has been calculated by Langer and Turski
\cite{LT73,TL80} and by Kawasaki \cite{Kawa}
for a liquid-gas phase transition near the critical point, where
the gas is not dilute, to be
\begin{equation}
\kappa = \frac{2\lambda\sigma T}{\ell^2 n_{\ell}^2 R_*^3} \, .
\end{equation}
This involves the thermal conductivity $\lambda$, the surface free energy
$\sigma$, the latent heat per molecule $\ell$ and the density of molecules
in the liquid phase $n_{\ell}$.
The interesting physics in this expression is the appearance
of the thermal conductivity.  In order for the droplet to grow beyond
the critical size, latent heat must be conducted away from the surface into
the gas.  For a relativistic system of particles or quantum fields which
has no net conserved charge, such as baryon number, the thermal conductivity
vanishes.  The reason is that there is no rest frame defined by the baryon
density to refer to heat transport.  Hence this formula obviously cannot
be applied to such systems.  The dynamical prefactor for such systems was first 
evaluated a' la Langer by two of us \cite{CK92D} to be
\begin{equation}
\kappa = \frac{4\sigma}{(\Delta w)^2 R_*^3} \left[ \frac{4}{3}
\eta + \zeta \right]\, .
\end{equation}
Here $\eta$ and $\zeta$ are the shear and bulk viscosities, respectively, and 
$\Delta w$ is the enthalpy density difference between the two phases (equal to 
the latent heat since the pressures are equal at the critical temperature).  
This fully relativistic expression was subsequently generalized by Venugopalan 
and Vischer \cite{VV} to include a conserved baryon number.
\begin{equation}
\kappa = \frac{2\sigma}{(\Delta w)^2 R_*^3} \left[ \lambda T +
2 \left( \frac{4}{3}\eta + \zeta \right) \right]
\end{equation}
In the non-relativistic limit and when the viscosities are small compared to the 
heat conductivity this reduces to Langer's expression.  In the relativistic 
limit with no net baryon number, where effectively $\lambda \rightarrow 0$, this 
reduces to equation (21).

Unfortunately there are several erroneous expressions for $\kappa$ which 
subsequently appeared in the literature.  In \cite{RF96} one finds that in the 
relativistic limit with no net baryon number
\begin{equation}
\kappa = \sqrt{\frac{2 \sigma w_h}{(\Delta w)^2 R_*^3}} \, .
\end{equation}
This was "derived" under the assumption of vanishing shear viscosities.  An 
expression for $\kappa$ in the nonrelativistic limit with non-zero baryon number 
was found which differed from Langer's result too.  By including the viscosities 
another paper \cite{SMG00} finds that
\begin{equation}
\kappa = \sqrt{\frac{2 \sigma w_h}{(\Delta w)^2 R_*^3}}
+ \frac{1}{c_s^2}  \frac{\sigma}{(\Delta w)^2 R_*^3} \left[ \frac{4}{3}
\eta + \zeta \right]\, .
\end{equation}
Here $c_s$ is the speed of sound in the low density phase.
In the limit that the viscosities vanish this reduces to the result of 
\cite{RF96}, and in the limit that they are large this reduces to the result of 
\cite{CK92D} albeit with a factor of $1/c_s^2$ rather than 4.

The results found by \cite{RF96} and \cite{SMG00} are suspicious because they 
predict a dynamical growth factor even in the absence of viscous forces.  These 
results are wrong because of a misinterpretation of a derivative.  Both papers 
follow the approach of \cite{CK92D} relatively closely except for one crucial 
point.  At some point in the analysis one encounters the derivative
$\partial f(\overline{e},T)/\partial \overline{e}$ evaluated at 
the radial profile $\overline{e} = \overline{e}(r)$ of the energy density
for a spherical bubble or droplet connecting the two phase.  Thus 
$\overline{e}(r)$ varies between $e_l$ and $e_h$.  The free energy density 
varies accordingly.  As explained carefully and in some
detail in \cite{CK92D}, this derivative is taken at fixed 
temperature.  At the equilibrium points it is zero: see the discussion in 
section 2 above.  However, both \cite{RF96} and \cite{SMG00} equate $f$ with
$-p$ irrespective of whether we are at one of the minima or not.  They then 
assume an equation of state $p = c_s^2 e$ to finally obtain 
$\partial f(e,T)/\partial e = -c_s^2$.  From that point on the analysis 
diverges, and the erroneous results for $\kappa$ inevitably follow.

It is also possible to argue that the erroneous results for $\kappa$ are not 
only mathematically incorrect but physically incorrect too.
What is relevant is the slowest required process, not the fastest.
Furthermore it is incorrect to add up the
rates of all processes to obtain the highest possible phase transition
speed.  If not all required processes are completed the phase transition
is not complete either. For example to establish pressure (momentum)
equilibrium \cite{RF96,SMG00} is not sufficient, because then a subsequent
step of establishing thermal equilibrium is required, 
and only when both processes are over has the phase transition been completed.

In ref. \cite{CK92D} the slowest process, heat transfer via
viscosity, was evaluated. In some cases, when we have non-negligible net
baryon charge in our system, the heat conductivity may also contribute to
the heat transfer to the new phase, thus speeding up somewhat this
slowest of all processes \cite{VV}. If the transport coefficients are
all vanishing so that thermal balance cannot be achieved, we will never
reach both phase and thermal equilibrium, so the rate tends to zero
as stated in ref. \cite{CK92D}.

In case other degrees of freedom exist which permit energy transport
between the phases leading to a common temperature \cite{I97,AB99}, these 
processes can be combined with other transport processes like viscosity
and/or heat conduction leading to faster temperature balance and a higher rate.
However, sound waves do not lead to dissipation as perfect fluid dynamics is 
adiabatic (in the absence of shock waves). Thus, the sound wave as the sole 
mechanism cannot characterize dissipative transport processes, as was mistakenly 
claimed in \cite{RF96} and \cite{SMG00}.

A final remark: If the external dynamical evolution is much faster than
the processes involved in the phase and kinetic equilibration then the
matter involved in the phase transition loses both phase and thermal 
equilibrium, and we have to abandon
the thermal and fluid dynamical approaches altogether.  An example 
in the context of high energy heavy ion collisions is presented in \cite{CM95} 
where an effective field theoretical approach is used.
Generally this will be the case for small, rapidly developing systems only.

\section{Conclusions}

In this paper we have examined two subtle issues in the dynamics of domain 
walls.  One of them is the difference between the effective potential as a 
function of a scalar field and the free energy as a function of energy density, 
both evaluated near the critical temperature.  The other involves a 
misconception in the literature about the derivation of the dynamical prefactor, 
or growth rate, in homogeneous nucleation theory.  It is important to understand 
these subtleties because of their importance in such physical environments as 
cosmology, astrophysics and high energy nuclear collisions.

\section*{Acknowledgements}

This work was supported by the US Department of Energy under grant
DE-FG02-87ER40328 and by the Research Council of Norway.

\newpage

\begin{figure}[h]
\centerline{\epsfig{figure=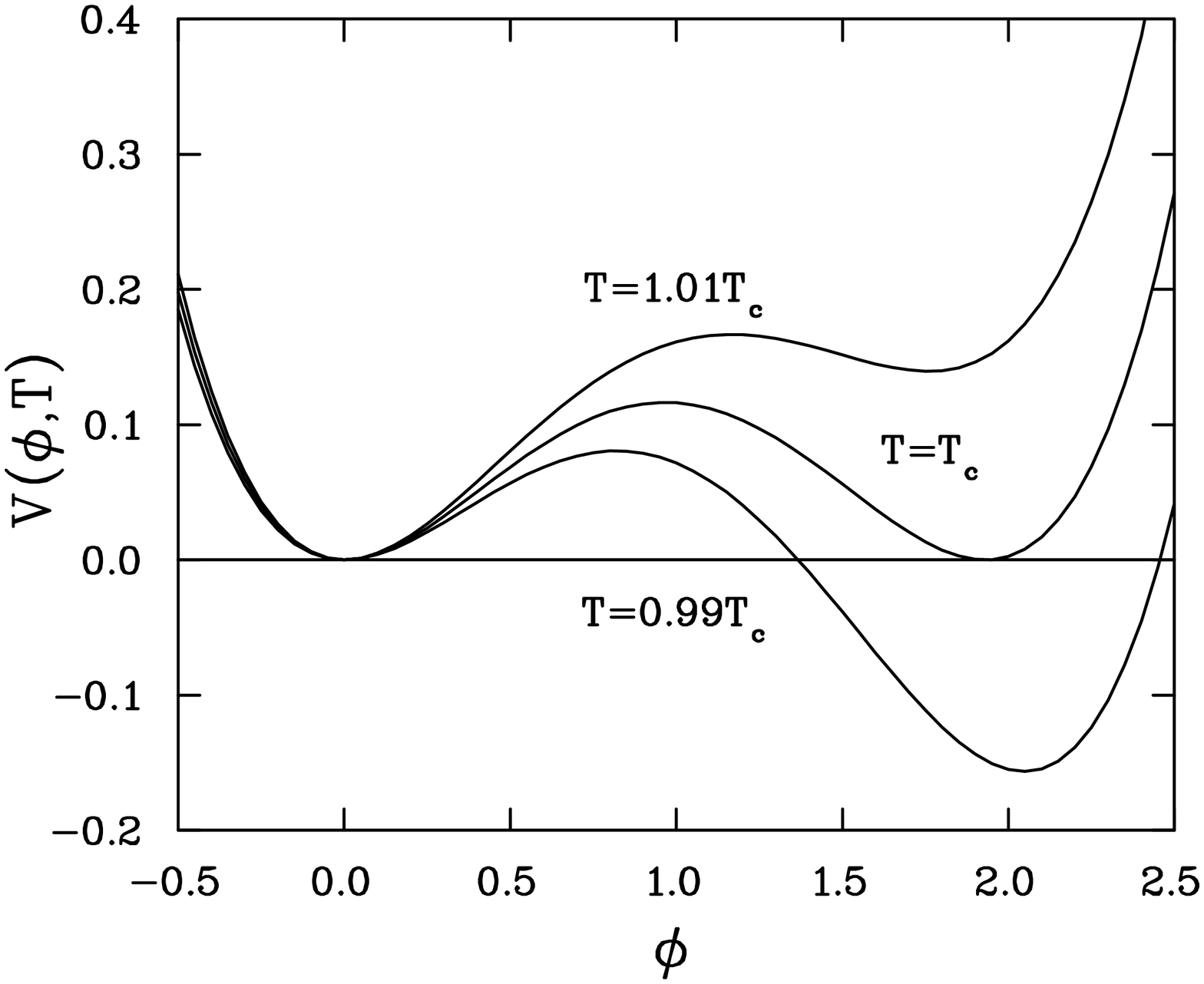,width=12.0cm,angle=90}}
\caption{The effective potential as a function of the field.}
\end{figure}

\begin{figure}
\centerline{\epsfig{figure=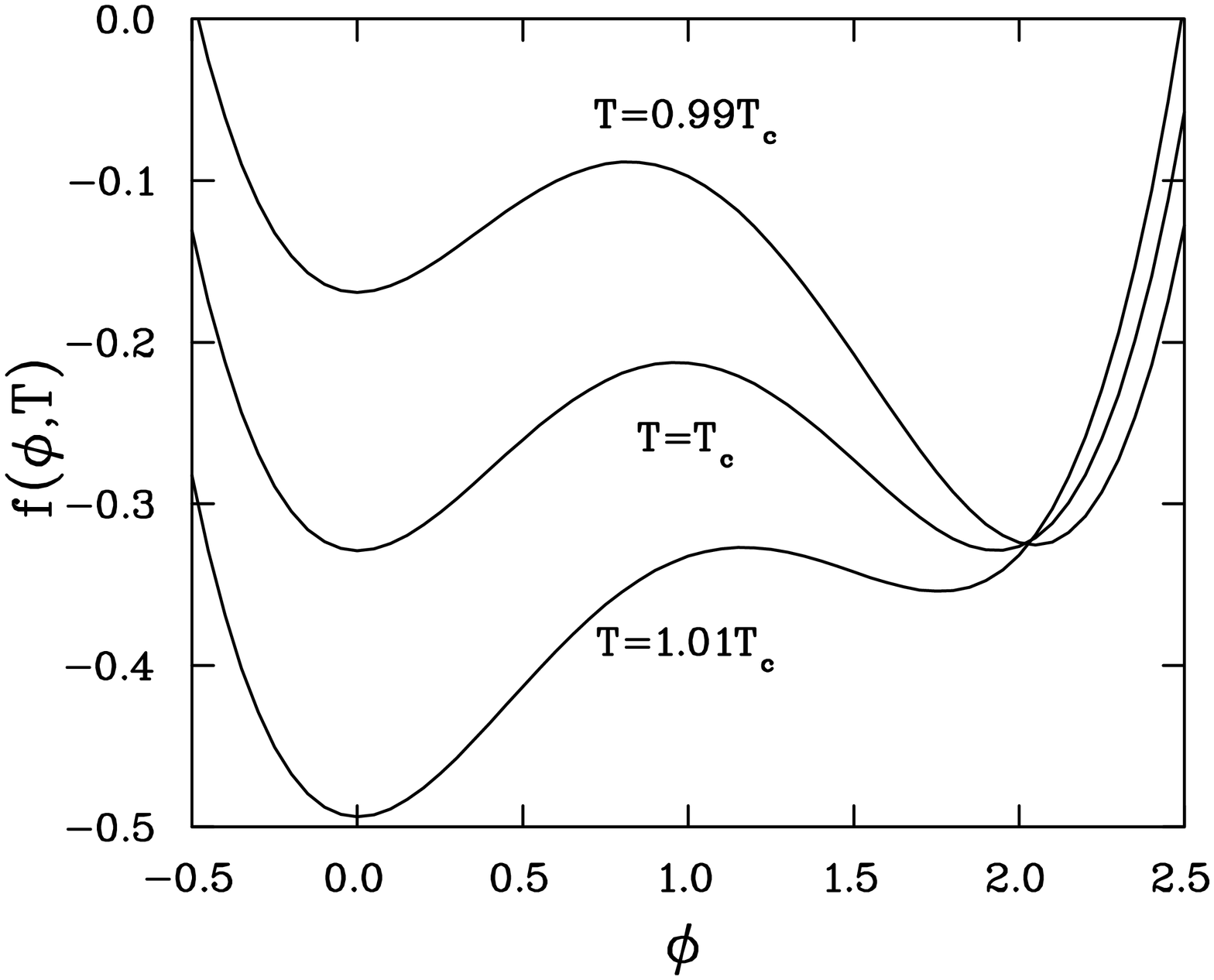,width=12.0cm,angle=90}}
\caption{The free energy density as a function of the field.}
\end{figure}

\begin{figure}[t]
\centerline{\epsfig{figure=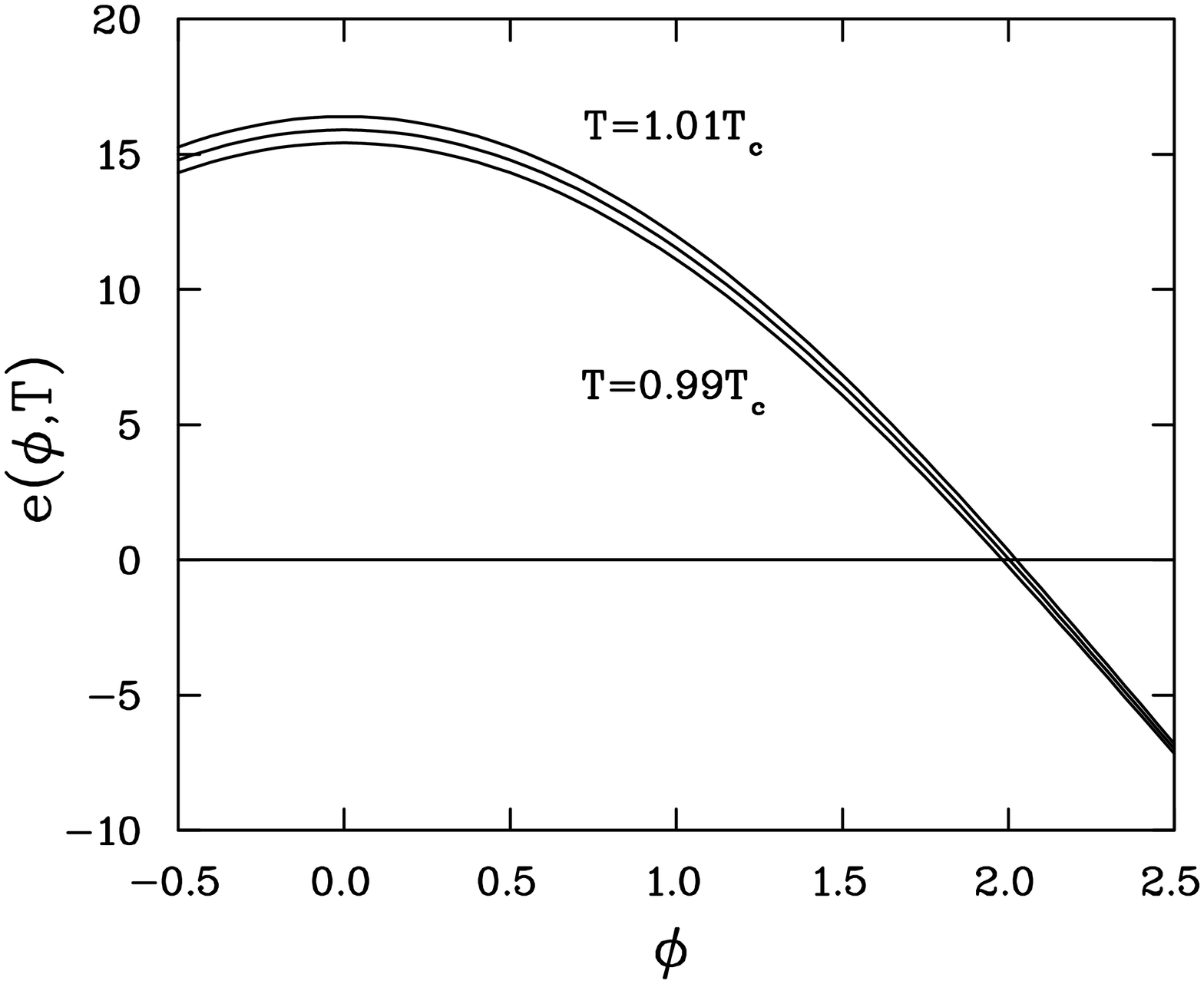,width=12.0cm,angle=90}}
\caption{The energy density as a function of the field.}
\end{figure}

\begin{figure}[b]
\centerline{\epsfig{figure=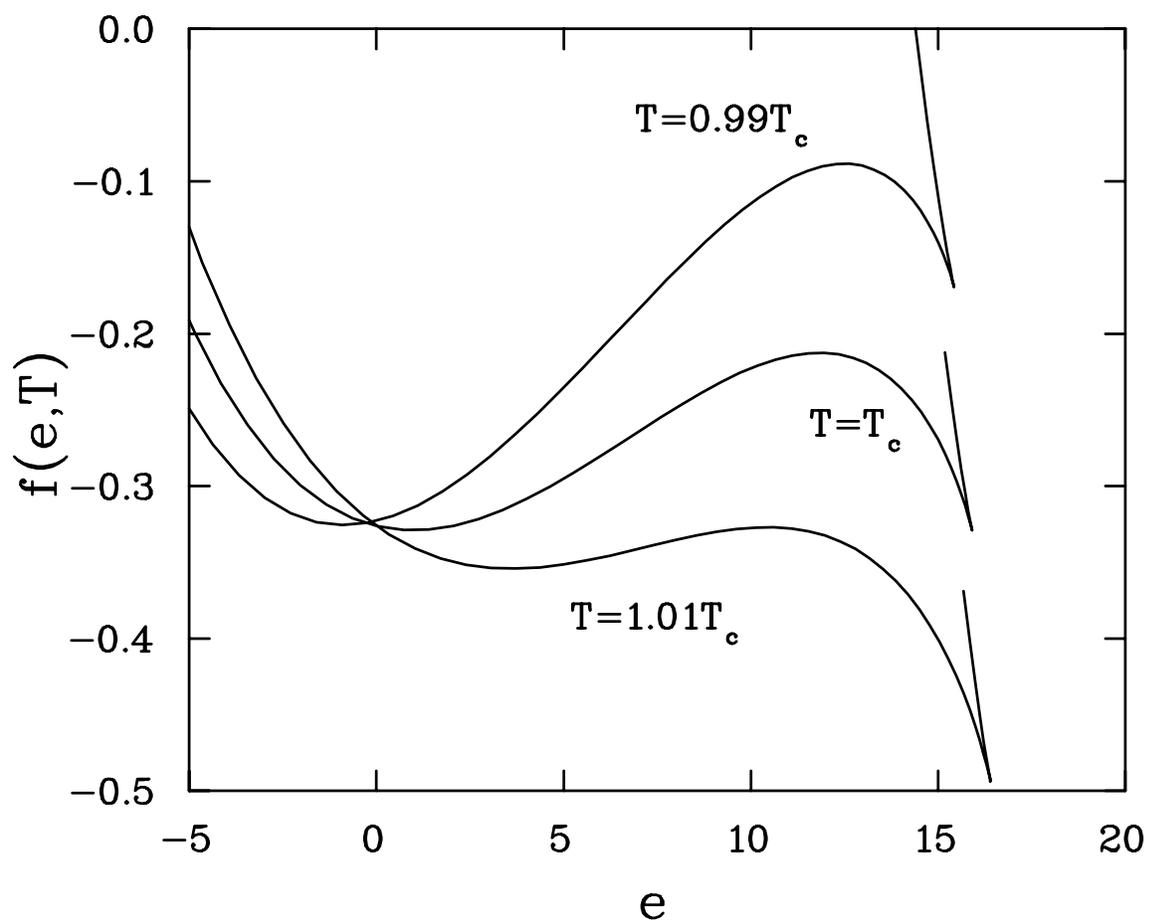,width=12.0cm,angle=90}}
\caption{The free energy as a function of the energy density as
derived from figures 1-3.}
\end{figure}

\begin{figure}
\centerline{\epsfig{figure=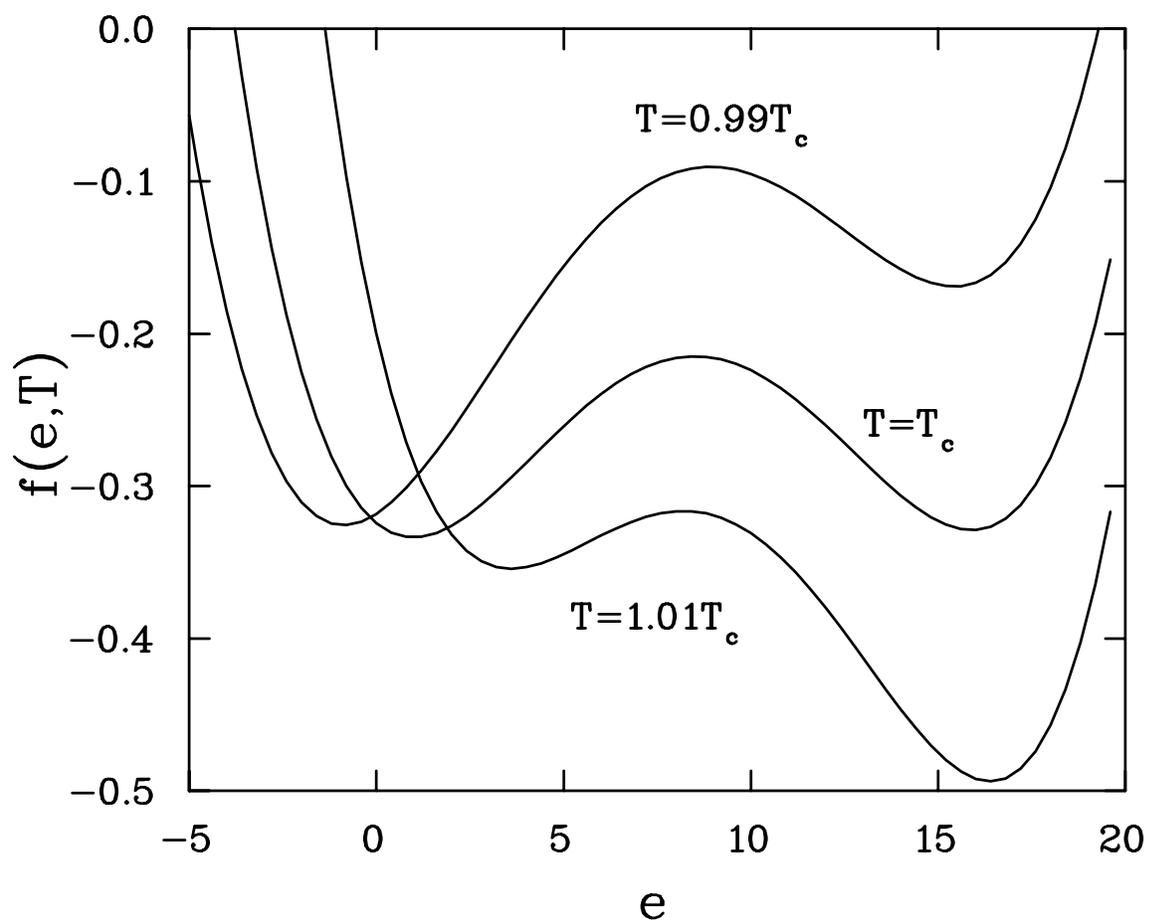,width=12.0cm,angle=90}}
\caption{The free energy as a function of the energy density
from a Landau construction.}
\end{figure}

\end{document}